# Forensic Investigation of P2P Cloud Storage: BitTorrent Sync as a Case Study

Teing Yee Yang[1], Ali Dehghantanha[2], Kim-Kwang Raymond Choo[3], Zaiton Muda[1]

1 Department of Computer Science, Faculty of Computer Science and Information Technology, Universiti Putra Malaysia, UPM Serdang, Selangor, Malaysia

2 The School of Computing, Science & Engineering, Newton Building, University of Salford, Salford, Greater Manchester, United Kingdom

3 Information Assurance Research Group, University of South Australia, Adelaide, South Australia, Australia.

## Abstract

Cloud computing has been regarded as the technology enabler for the Internet of Things (IoT). To ensure the most effective collection of IoT-based evidence, it is vital for forensic practitioners to possess a contemporary understanding of the artefacts from different cloud services. In this paper, we seek to determine the data remnants from the use of BitTorrent Sync version 2.0. Findings from our research using mobile and computer devices running Windows 8.1, Mac OS X Mavericks 10.9.5, Ubuntu 14.04.1 LTS, iOS 7.1.2, and Android KitKat 4.4.4 suggested that artefacts relating to the installation, uninstallation, log-in, log-off, and file synchronisation could be recovered, which are potential sources of IoT forensics. We also present a forensically sound investigation methodology for BitTorrent Sync.

## 1. Introduction

The Internet of things (IoT) has been the focus of researchers and practitioners in recent years, due to the increasing popularity of internet connected devices. Gartner (2014a) forecasted the number of IoT devices to reach 26 billion by 2019. Similarly, the International Data Corporation (IDC) (2014) predicted that the IoT devices to hit 30 billion by 2020, amounting to USD3.04 trillion. Since the IoT devices are equipped with low storage and computational capability (Zawoad, 2015), the IDC (2014a) predicted that 90% of all IoT data will be hosted on cloud service provider platforms by 2019 as cloud computing reduces the complexity of supporting IoT data blending.

Although cloud computing is often being credited for enabling promising and cost-competitive storage solutions for the IoT, it is subject to potential abuse by both traditional and cyber miscreants in the meantime (Choo, 2008). Potential crimes related to cloud computing include information theft (Choo, 2010; Symantec, 2011; Duke, 2014), malicious software distribution (Shado, 2014), denial of service attacks (DDoS) (Lemos, 2010; Peterson, 2013), industrial espionage, copyright infringement, and storage of illegal materials (e.g. child exploitation materials, and terrorism materials).

Since a public cloud storage infrastructure may constitute cloud servers located in one or more data centers and jurisdictions, the forensic community is often subject to various legal challenges (Taylor et al., 2011; Chung et al 2012; Grispos, Storer, and Glisson, 2013; Hooper, Martini, and Choo, 2013; NIST, 2014; Quick et al., 2014a; Martini and Choo 2014a). Even in the event that the evidence could be identified, it would not be trivial to seize the storage media (server) as it is likely to hold data belonging to other users (e.g. in a multi-tenancy cloud environment) (ENISA, 2012).

Due to the rapid advancement of the IoT, it is imperative that forensic examiners are cognisant of the different types of cloud products as well as an up-to-date understanding of the potential artefacts that could potentially be recovered to inform the IoT investigations (Hale, 2013; Quick and Choo, 2013a, 2013b, 2014; Martini and Choo, 2014c; Quick et al., 2014). Depending on the cloud storage solution in use, the client device can often provide potential for alternative methods for recovery of the cloud artefacts (Farina et al., 2014; Scanlon et al., 2014a; Scanlon et al., 2014b; Scanlon et al., 2015). Hence, in this paper, we seek to identify potential terrestrial artefacts that may remain after the use of the newer BitTorrent Sync version 2.0. Similar to the approaches of Quick and Choo (2013a, 2013b, 2014), we attempt to answer the following questions in this research:

1. Does the act of file download or file upload using BitTorrent Sync cloud storage alter the file contents and timestamps of the original files?
2. What data can be found on a computer hard drive and memory after a user has used the BitTorrent Sync client application and web application, and the location of data remnants on Windows 8.1, Ubuntu 14.04.1 LTS (Ubuntu), and Mac OS X Mavericks 10.9.5 (Mac OS)?
3. What data remains on an Apple iPhone 4 running iOS version 7.1.2 (iOS) and HTC One X running Android Kit Kat 4.4.4 (Android) mobile devices after a user has used the BitTorrent Sync client apps?
4. What data can be seen in network traffic?

We regard the contribution of this paper to be two-fold:

1. To provide the forensic community an in-depth understanding of the types of terrestrial artefacts that are likely to remain after the use of the newer BitTorrent Sync cloud applications.
2. An operational methodology to guide forensic practitioners in examining the latest BitTorrent Sync applications.

This paper is organised as follows. The background is provided in Section 2. In Section 3, we provide an overview of the research methodology including the cloud investigative framework used and the experiment setup. In Section 4 and 5, we detail the findings from the technical experiments involving the personal computers and mobile devices. Section 6 discusses the network artefacts. We outline our proposed operational methodology for BitTorrent Sync forensics in Section 7. Finally, we conclude the paper and outline potential future research topics in Section 8.

## 2. Background

BitTorrent Sync is a product built by BitTorrent Inc., the creator of the BitTorrent P2P file sharing protocol (BitTorrent Inc., 2015b). BitTorrent Sync allows file hosting and sharing across multiple platforms such as Windows, Mac OS, Linux, Android, iOS, and Windows mobile OS (BitTorrent, 2014). Other than the free space on a sync device, the fully p2p-based architecture does not limit the amount of data that can be synced. Hence, it is not surprising that it becomes a popular choice for file replication and synchronisation. For example, in less than two years after its release, the number of BitTorrent Sync users is reportedly over 10 million and the application had transferred over 80 Petabytes of data as of August 2014 (Pounds, 2014). The users are required to install the client application to use the service. For the Linux client application, the service is only accessible through ‘http://localhost:8888’ and the web interface is password protected.

### 2.1 Device and folder sharing

BitTorrent Sync users are required to create a private identity for the first device running BitTorrent Sync 2.0 or later. The identity holds the user name (provided by the user), device name, identity-specific certificate fingerprint, and a 33-digit key created using a public key infrastructure. The key contains the folder permissions used to establish connections with other devices and for licensing purposes (BitTorrent Inc., 2015f). Only a private identity is required for all connected/linked devices.

When a user adds a new folder to BitTorrent Sync, the user automatically becomes the owner of that folder. Other than devices sharing the same private identity, BitTorrent Sync permits the users to share folders with a device that has a different identity. The folder sharing is facilitated by a sync link (introduced in BitTorrent Sync version 1.4), which can be shared as a HTTPS link or QR code and contains the following details:

- Shared folder name (prefixed with ‘f=’),
- an approximation of the folder size in exponential format (prefixed with ‘sz=’),
- a 20-byte folder ID in base32 encoding (prefixed with ‘s=’),
- a temporary key (formerly known as ‘one-time secret’) in base32 encoding (prefixed with ‘i=’),
- link expiration time (prefixed with ‘e=’),
- and a base32-encoded peer ID (prefixed with ‘p=’) used to identify which of 2 peers is going to take a role of a server during key negotiation (BitTorrent Inc., 2015d).

BitTorrent Sync allows the users to configure the expiry time for the sync link (three days by default), a limit on the number of uses of the sync link, as well as the authorisation settings during folder sharing (BitTorrent Inc., 2015e).

When a sync link is shared from host to guest, the guest responds the host with the temporary key (contained within the sync link shared by the host) followed by a request containing the locally generated X.509 certificate (which contains the public key fingerprint, owner's name, and device name), allowing the host to validate the identity of the requester (guest) before sending the actual master key.

A standard master key constitutes capital letters from A to Z and numbers from 2 to 7; the first letter of a key represents its type (See Table 1) while the remaining is a 20-byte sequence (usually 32 symbols) in base32 format, with the exception of the ‘Read-Only’ (RO) key which is twice as long (65 symbols), holding an additional data encryption/decryption key (BitTorrent Inc., 2015c). Folders shared with a ‘Read & Write’ (RW) permission provide the peers the ability to add new files/folders or modify/remove the existing ones, and synchronise the changes to all the peer devices. On the other hand, modifications made to the shared folders with a RO permission will not be synchronised to other corresponding devices. All the BitTorrent Sync keys are generated using ED25519 and SHA3 cryptographic algorithms (BitTorrent Inc., 2015a).

Table 1: Key types (BitTorrent Inc., 2015c).

| Key type | Uses |
|---|---|
| **A** | Standard key with read-write permission. |
| **B** | Read-only key derived from the ‘A’-type key. |
| **C** | Time-limited read-only one-time key derived from the ‘A’ or ‘B’-type key. |
| **D** | Standard key with read-write permission capable to seed data to encrypted nodes. |
| **E** | Read-only key derived from the ‘D’-type key which provides the ability to get and decrypt data from encrypted nodes (nodes |
| | with 'D' and 'F' types of key). |
| **F** | Encrypted key derived from the 'E' or 'F'-type key which provides the ability to receive, store and seed data, but cannot decrypt filenames or content. |
| **R** | Obsolete Read-only key generated by pre-1.0 version. This key is still in use for compatibility purposes. |

## 2.2 Data transfers

To initiate download of a shared folder, the guest must first download the metainfo (.TORRENT) file from the host. The guest then interprets the metainfo file for the folder metadata as well as tracker URL (or IP address and port combinations), and uses the details to locate peers actively participating in the particular swarm[1] or sharing the same share ID[2] through one or more methods as outlined in Table 2 (BitTorrent Inc., 2008; BitTorrent Inc., 2015a). Once peers have established a connection, they exchange peer lists to augment the peer list supplied by the tracker through a process known as peer exchange (Hunt, 2014). The data transfers are facilitated by the BitTorrent protocol, which uses a combination of TCP/IP and Micro Transport Protocol (µTP) as its transport protocol. All traffic between devices is encrypted with AES-128 in counter mode, using a unique session key derived from the RO key (BitTorrent Inc., 2015b).

Table 2: Peer discovery methods.

| Peer discovery method | Description |
|---|---|
| **Use relay server when required** | This option allows BitTorrent Sync to use the BitTorrent's relay server as an intermediary for connection with peers in the scenario when direct connection is not possible as a result of sophisticated NATs, firewalls, proxy servers, etc. |
| **Use tracker server** | A tracker server holds a list of share IDs as well as network information (e.g., IP addresses and port numbers) needed for peers establish direct connections with other peers. Peers with internal subnet matches will be directed to connect using LAN. |
| **Search Local Area Network (LAN)** | When LAN discovery is enabled, BitTorrent Sync sends multicast packets (which contains the requesting share IDs as well as IP address and port number combinations of the requesting nodes) across the local network to locate nodes sharing the same share IDs. |
| **Search Distributed Hash Table (DHT) network** | This option enables BitTorrent Sync to connect to a BitTorrent/uTorrent's DHT network and subsequently uses the DHT table to locate peers that share the same share IDs, eliminating the need for a tracker server. |
| **Predefined hosts** | This option allows peers to be contacted directly through a list of explicitly defined IP address and port combinations. |

During synchronisation, each data is broken down into small pieces prior to transmission. In order to achieve minimal bandwidth usage, only pieces containing changes are transmitted. The folder contents are compared between all peers from time to time to determine changes and the sync files are replaced by the newest version held by any peer (BitTorrent Inc., 2015b). The overwritten or deleted files will be kept in the folder's archive for 30 days by default, but the duration can be modified by the user.

## 2.3 Related Work

Scholars have noted the legal challenges involving cross-jurisdictional cloud forensic investigations (Mason and George, 2011; Taylor et al, 2012; Hooper et al., 2013), as well as the complexity and research opportunities in relation to cloud forensic investigations (Dykstra and Sherman, 2011; Birk et al., 2011; Ruan et al. 2011, Dominik, 2011; Damshenas et al., 2012; Daryabar et al., 2013; Simou, 2014). Research with a technical focus generally aims to address particular challenges associated with the on-device and remote collections of data artefacts from a decentralised cloud infrastructure (Zafarullah et al., 2011; Marty, 2011; Martini and Choo, 2012; Martini and Choo, 2013; Quick et al., 2014; Dykstra and Sherman, 2013; Zawoad et al., 2013; Gebhardt and Reiser, 2013; Martini and Choo 2014c). The studies of Chung et al. (2012), Quick and Choo (2013a, 2013b, 2014), Hale (2013), Thethi and Keana (2014), Oestreicher (2014), Farina et al. (2014), Shariati, Dehghantanha, and Choo (2015), Shariati, Dehghantanha, Martini, and Choo (2015), and Martini, Do, and Choo (2015) found that metadata and other data artefacts could potentially be recovered from client devices used to access cloud services, even when the data has been erased using eraser software, such as Eraser and CCleaner (Quick and Choo, 2013a, 2013b, 2014). Quick and Choo (2013c) also determined that the act of downloading data from cloud counterparts using a web or client application does not modify the hash of the data of relevance despite changes in file creation/modification times. The effectiveness of commercial forensic tools (e.g. Guidance EnCase, the Forensics Tool Kit (FTK), Memoryze, and AWS Export) in acquiring evidence remotely from the Amazon EC2 servers has also been studied (Dykstra and Sherman, 2012).

[1] A swarm a collection of peers sharing the same torrent file (BitTorrent Inc., 2014).

[2] A share ID is the SHA1 of the master key used by a share folder (Scanlon et al., 2014a).

A number of cloud-focused forensic frameworks and investigative guidelines have also been proposed in the literature. The first cloud forensic framework was proposed by Martini and Choo (2012), which was used to investigate ownCloud (Martini and Choo, 2013), Amazon EC2 (Thethi and Keana, 2014), VMWare (Martini and Choo 2014c), XtreemFS (Martini and Choo 2014b), Ubuntu One (Shariati, Dehghantanha, Martini and Choo 2015), SugarSync (Shariati, Dehghantanha and Choo 2015), etc. The four-stage framework was subsequently extended and validated using SkyDrive, Dropbox, Google Drive and ownCloud (Quick et al. 2014). Chung et al. (2012) proposed a cloud investigation guideline, which was used to investigate Amazon S3, Google Docs, and Evernote on Windows, Mac OS, iOS, and Android devices.

In to the context of our study, Farina et al. (2014) examined the potential of recovering forensic artefacts from computers (running Windows XP, Windows 7, and Linux Debian) and network capture involving the use of BitTorrent Sync (version 1.1.82), and Scanlon et al. (2014a, 2014b, 2015) presented an investigative framework for the remote collection of evidence from a decentralised file synchronisation network. Since a redesigned folder sharing workflow has been introduced in the newer version of BitTorrent Sync (from version 1.4 onwards; BitTorrent, 2015d), there is a need to develop an up-to-date understanding of the artefacts from the newer BitTorrent Sync applications.

## 3. Research Methodology

In this research, we adopt the cloud investigative framework proposed by Martini and Choo (2012) – see Figure 1. As explained by Martini and Choo (2012), a key characteristic of their framework is the iteration introduced in the Examination and Analysis phase. This allows one or more simultaneous iteration(s) of the framework with evidence source identification and preservation via the associated parties (i.e., CSP, peer node users when undertaking p2p storage cloud investigation, etc.) when evidence of cloud computing use is subsequently discovered on a client device. We demonstrate the utility in the context of this research as follows:

1. *Evidence source identification and preservation.* In the first phase, the physical hardware of interest was identified, which contained the virtual disk files (VMDK) and virtual memory files (VMEM) in each VM folder. The mobile devices used in this research were a HTC One X running Android KitKat 4.4.4 and an Apple iPhone 4 running on iOS version 7.1.2. A forensic copy was created for each VMDK and VMEM file in E01 container and raw image file (.dd) formats respectively. For the mobile devices, we made a bit-for-bit image of the internal storage and subsequently converted the images to the E01 container format. An MD5 and SHA1 hash was calculated for each original file and subsequently verified for each copy.
2. *Collection*. In this phase, we collected data containing the details needed for analysis from the forensic images (see Section 4). Similar to the earlier evidence source identification and preservation phase, an MD5 and SHA1 hash was calculated for each original file and subsequently verified for each collected or exported file.
3. *Examination and Analysis.* This phase is concerned with examination and analysis of data at rest, in motion, or in execution collected in our research. The search terms were determined from the filenames observed, text from within the Enron data files, as well the BitTorrent Sync instances created during the research. These included:
    - 'bittorrent', 'BitTorrent Sync', 'torrent', 'btsync'
    - 'enron3111', '3111', 'Enron'
    - identity names such as 'host' and 'guest'
    - Sync links, share IDs, peer IDs, folder IDs, certificate fingerprints, temporary Keys, and other IDs and keys relevant to BitTorrent Sync.

   In this research, we started by analysing the guest VMs/devices. Afterwards, we iterated the framework with evidence source identification, preservation, and analysis via the host/peer VMware Workstations (VMs) using the BitTorrent Sync artefacts recovered from the guest VMs or devices.
4. *Reporting and presentation.* We reported our findings, as described in this paper.

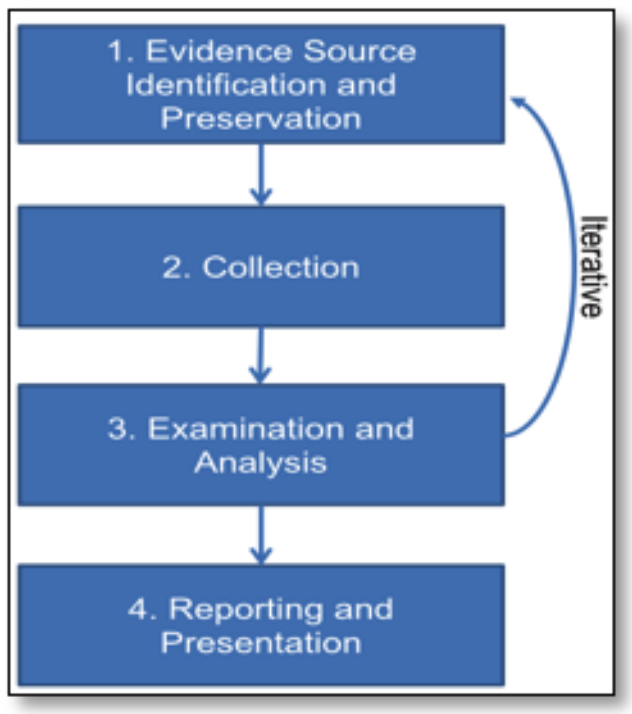

Figure 1: Cloud forensics framework of Martini and Choo (2012).

## 3.1 Experimental Setup

Two VMs were created for each operating system (OS) investigated to represent the host and the guest workstations. As explained by Quick and Choo (2013a, 2013b, 2014), using physical hardware to undertake setup, erasing, copying, and re-installing would have been an onerous exercise. Moreover, a virtual machine allows room for error by enabling the test environment to be reverted to a restore point if the results are unfavourable. The hard drive and RAM were configured with minimal space in order to reduce the time required to analyse the considerable amounts of snapshots. A total of 24 VM snapshots was made of each workstation representing 24 real life scenarios of using BitTorrent Sync (e.g., install, access, upload, download, view, delete, and uninstall) on various operating systems - see Table 3. For the purpose of computer forensic analysis, the data sharing was only limited to the default peer discovery setting (by having the 'Use relay server when required', 'Use tracker server', and 'Search LAN' options checked) with a Read and Write permission.

Table 3: Configurations of virtual machines for BitTorrent Sync client application analysis on Windows 8.1.

| OS | Host VM/Guest VM | VM details |
|---|---|---|
| **Windows 8.1 (client application)** | **Base-VM**<br>1.0, 2.0, 3.0 | A base VM snapshot was prepared for each OS as a control media to determine changes during each experiment with the following configurations:<br>• Windows 8.1 Professional (ServicePack 1, 64-bit, build 9600) with 2GB RAM and 20GB hard disk (1.0).<br>• Ubuntu 14.04.1 LTS with 1GB RAM and 20GB hard disk (2.0).<br>• Mac OS X Mavericks 10.9.5 with 1GB RAM and 60GB hard disk (3.0). |
| | **Install-VM**<br>1.1, 2.1, 3.1 | By duplicating a copy of the base snapshot (1.0, 2.0, and 3.0), we accessed the BitTorrent Sync website (*https://www.getsync.com/*) to download and subsequently install BitTorrent Sync version 2.0.93 (the latest version at the time of this research. A separate identity was created for each device/VM. |
| | **Access-VM**<br>1.1.1, 2.1.1, 3.1.1 | A copy of install snapshot (1.1, 2.1, and 3.1) was made to examine the process of logging in the BitTorrent Sync client application. |
| | **Upload/Download-VM**<br>1.1.2, 2.1.2, 3.1.2 (Synchronise) | A second copy of the install snapshot (1.1, 2.1, and 3.1) was made to examine the process of syncing files using the default peer discovery settings. The Enron dataset files were copied from the host machine to *C:\Sync*, */home/[User Profile]/Sync/*, and */Users/[User Profile]/Sync/* of the Windows, Ubuntu, and Mac OS host workstations. A sync link was then generated for the sync directory and subsequently used to link with the guest workstaion. The creation, modified, and last accessed times of each file were noted to detect changes in timestamps after transferring files. |
| | **Delete-VM**<br>1.1.2.1, 2.1.2.1. 3.1.2.1 (Synchronise) | A copy of the upload/download snapshot (1.1.2, 2.1.2, 3.1.2) was created to assess the process of deleting the uploaded files on the host workstations and determined changes to the guest workstations and vice versa. |
| | **Disconnect-VM**<br>1.1.2.2, 2.1.2.2, 3.1.2.2 | A second copy of the upload/download snapshot (1.1.2, 2.1.2, 3.1.2) was made to examine the process of disconnecting a shared folder. The option 'Delete files from this device' option was selected to remove the synced files completely from the host workstations. |
| | **Uninstall-VM/**<br>1.1.2.3, 2.1.2.3, 3.1.2.3 | A final copy of the upload/download snapshot (3.1.1) was made to examine the process of uninstalling the BitTorrent Sync client application. Since BitTorrent Sync does not come with an uninstaller, the uninstallation was undertaken using the Windows '*Programs and Features*' function in the Control Panel; the commands "find / -name ".sync" -type d -exec rm -rf {} \" and "find / -name "BitTorrent Sync" -type f -exec rm -rf {} \" on the Ubuntu OS workstation. manual dragging of the BitTorrent Sync folders of relevance to the Trash directory on the Mac OS workstation. |
| | **Unlink-VM** 1.1.3, 2.1.3, 3.1.3 | A final copy of the install snapshots (1.1, 2.1, and 3.1) was made to investigate the process of unlinking an identity on the desktop clients investigated. |

Similar to the approaches of Quick and Choo (2013a, 2013b, 2014) and Shariati et al. (2015a, 2015b), the 3111th email messages of the UC Berkeley Enron email dataset (downloaded from *http://bailando.sims.berkeley.edu/enron_email.html* on 24th of September 2014) were used to create the sample files and saved in .RTF, .TXT, .DOCX, .JPG (print screen), .ZIP, and .PDF formats. Each VM was shut down and a snapshot was taken of the VM after each experiment occurred, allowing the VM to be reverted back to this state when needed. The RAM captures were taken immediately after each experiment, just prior to shutdown in our research. The physical memory dumps were instantiated by the Virtual Memory (.VMEM) files (created by VMware) to represent captures of memory dumps which are not being adulterated with the use of memory acquisition software (Quick and Choo, 2013a; 2013b). A similar consideration was made with respect to running/hosting physical acquisition and network capture software on the VMs. Hence, we instantiated the physical hard drive with the Virtual Machine Disk (.VMDK) files (created by VMware) and hosted the packet capture software on the local host.

In order to undertake analysis into the mobile apps, we prepared a default factory restored iPhone 4 running iOS 7.1.2 and a HTC One X running Android KitKat 4.4.4. We then jailbroke/rooted both the devices using Pangu8 v1.1 and Odin3 v.185 to enable root access, respectively. To examine the matter in which the file systems were treated in relation to different BitTorrent Sync usage scenarios, we created a series of physical images of the mobile devices using dd over SSH/ADB Shell. In particular, the first image was undertaken prior to the installation of the BitTorrent Sync apps to create the control base images for this research. Then, the BitTorrent Sync iOS app version 2.0.27.1 and Android app version 2.0.85.0 were installed on the respective devices to make the second image respectively. A third image was undertaken after downloading sync files from the 'H1.1.1 Upload-VM' (see Table 1). Additionally, an extra image was made of the Android device to examine the process of adding and uploading a shared folder using the BitTorrent Sync app (unsupported by the iOS app). Next, we created an image of both the devices to assess the process of deleting the shared folder. The final image was made following the uninstallation of the apps.

The packet capture software was started prior and stopped immediately after each experiment was carried out. The experiments were predominantly undertaken in NATed (where NAT stands for Network Address Translation) network environment and without firewall outbound restriction to represent a typical BitTorrent Sync usage situation. Each experiment was repeated at least thrice (at different dates) for consistency of findings. Table 4 details the tools prepared for this research.

Table 4: Tools prepared.

| Tool | Usage |
|---|---|
| **FTK Imager Version 3.2.0.0** | To create forensic images for the .VMDK files. |
| **dd version 1.3.4-1** | To produce a bit-for-bit image of mobile devices' internal storage as well as .VEM files. |
| **Autopsy 3.1.1** | To parse the file system, produce directory listings, as well as extracting or analysing stored files, browsing history, 'NTUSER.dat' registry files (using the RegRipper plugin), 'pagefile.sys' Windows swap file, and unallocated spaces located within the forensics images of VMDK files. |
| **emf_decrypter.py** | To decrypt the iOS images acquired for analysis. |
| **HxD Version 1.7.7.0** | To conduct keyword searches in the unstructured datasets. |
| **Volatility 2.4** | To analyse the running processes (using the 'pslist' function), network statistics (using the 'netscan' function), and detecting the location of a string (using the 'yarascan' function) recorded in the physical memory dumps. |
| **SQLite Browser Version 3.4.0** | To view the contents of SQLite database files. |
| **Wireshark version 1.10.1** | To analyse the network traffic. |
| **Network Miner version 1.6.1** | To analyse and data carve the network files. |
| **Whois command** | To determine the registration information of the IP addresses. |
| **Photorec 7.0** | To data carve the unstructured datasets. |
| **File juicer 4.45** | To extract files from files. |
| **Nirsoft Web Browser Passview 1.19.1** | To recover the credential details stored within web browsers. |
| **Nirsoft cache viewer, ChromeCacheView 1.56, MozillaCacheView 1.62, IECacheView 1.53** | To analyse the web browsing cache. |
| **BrowsingHistoryView v.1.60** | To analyse the web browsing history. |
| **Thumbcacheviewer Version 1.0.2.7** | To examine the Windows thumbnail cache. |
| **Windows Event Viewer Version 1.0** | To view the Windows event logs. |
| **Console Version 10.10 (543)** | To view the Mac-OS-specific log files (e.g., Apple System Logs). |
| **Windows File Analyser 2.6.0.0** | To analyse the Windows prefetch and link files. |
| **Plist Explorer v1.0** | To examine the contents of the Apple PLIST files extracted from iPhone Analyser. |
| **chainbreaker.py** | To extract the master keys stored in Mac's Keychain dump. |
| **NTFS Log Tracker** | To parse and analyse the $LogFile, $MFT, and $UsnJrnl New Technology File System (NTFS) files. |
| **BEncode Editor v0.7.1.0** | To view the contents of bencode files. |

## 4. BitTorrent Sync analysis on desktop clients

Before undertaking the evidential analysis, we collected test data that matched the search terms 'Bittorent sync', 'btsync', and 'Enron' in the hard disk images, but held formats unsupported by the Autopsy forensic browser for analysis using the tools of relevance in the latter phase. These included SQLite database files, PLIST files, prefetch files, event logs, shortcuts, thumbnail cache, $MFT, $LogFile, $UsnJrnl, as well as web browsers' data folders/files (e.g., *%AppData% \Local\Google*, *%AppData% \Local\Microsoft \Windows \WebCache*, *%AppData% \Roaming\Mozilla*, *%AppData% \Local\Microsoft \Windows \Temporary Files \index.dat*). The volatile data was collected using the Volatility tools, Photorec file carver, and HxD Hex Editor for the physical memory dumps, and Wireshark and Netminer network analysis software for the network captures.

Whilst undertaking keyword search for the data of relevance, we determined that there was no data related to BitTorrent Sync and the Enron emails on the control base VM snapshots (1.0, 1.1 IE, 1.2 MF, 1.3 GC, 2.0, and 3.0). This suggested that the BitTorrent Sync/Enron related data located in the remaining snapshots were remnants from BitTorent Sync use. An inspection of the metadata of the downloaded files on the Windows 8.1 client observed that the last accessed and modified timestamps were the times when the files were downloaded, and the last written timestamps retained its original value unchanged. On the Ubuntu client, the added timestamps were the times when the files were downloaded, while all other timestamps (i.e., modification, creation, and last opened) remained unchanged. As for the Mac OS client, only the accessed timestamps matched the file download times; the modification timestamps preserved its original timestamps. In all cases, we determined that the MD5 and SHA1 hash values for the downloaded files were similar to the that of the original copies, suggesting that no alteration was made during the file transfers.

### 4.1 Directory listings and files of forensic interest

The downloaded folders were saved at *%Users%\[User Profile]\BitTorrent Sync*, */home/[User profile]/BitTorrent Sync*, and */Users/[User Profile]/BitTorrent Sync* on the Windows 8.1, Ubuntu OS, and Mac OS clients by default, respectively; Within the shared folders (both locally added and downloaded) there is a hidden '.sync' subfolder. The file of particular interest stored within the subfolder is the 'ID' file which holds the folder-specific share ID in hex dump format. The share ID would be especially useful when seeking to identify peers sharing the same folder during network analysis (Scanlon et al., 2014a, 2014b).

When a synced file was deleted, we were able to recover copies of the deleted file from the */.sync/Archive* folder of the corresponding peer devices. Depending on the duration configured by the users, it is important to note that the deleted files will only be kept in the archive folder for 30 days by default (BitTorrent Sync, 2015). The presence of the '.sync' subfolder also means when a user synchronises or deletes a file in BitTorrent Sync Windows application, there will be filename and timestamp references for the synced or deleted files in NTFS files such as $LogFile, $MFT, $UsnJrnl to identify its use.

In addition, we could recover copies of the deleted files alongside the pertinent file deletion information (e.g., the original paths, file sizes, and deletion times) from the *%$Recycle.Bin%\SID* folder on Windows 8.1, but the filenames were renamed to a set of random characters prefixed with $R and $I. On Ubuntu OS, we were able to recover copies of the deleted files from the */home/[User Profile]/.local/share/Trash/files* trash folder. To identify the original file paths as well as deletion times, we analysed the .TRASHINFO files located in */home/[User Profile]/.local/share/Trash/info/*. In contrast to Windows and Ubuntu OS, examination of the Mac OS trash folder (located at */Users/[User profile]/.Trash*) was only able to recover copies of the deleted files. However, it is

noteworthy that the findings are only applicable to the system that initiated the file deletion and in the circumstance when recycle bin or trash folder is not emptied.

Other than inspecting the directory listing, we identified that the practitioner could potentially recover the BitTorrent Sync usage information from various metadata files resided in the application folder located at *%AppData%\Roaming\BitTorrent Sync* on Windows 8.1 (was previously stored at *%Documents and Settings%\[User Profile]\Application Data\BitTorrent Sync*) and */Users/[User Profile/Library/Application Support/BitTorrent Sync* on Mac OS. Similar to the installation folder, the application folder of the Linux client application is the directory where the application package is unpacked. The application folder maintains a similar directory structure across multiple operating systems, and the */%BitTorrent Sync%/.SyncUser<Random number>* subfolder is an identity-specific application folder that will be synchronised across multiple devices sharing the same identity.

The first file of particular interest with the application folder is the settings.dat file. This file maintains a list of metadata associated with the device under investigation such as the installation path (which could be distinguished from the 'exe_path' entry), installation time in Unix epoch format ('install_time'), non-encoded peer ID ('peer_id'), log size ('log_size'), registered URLs for peer search ('search_list', 'tracker_last' etc.), and other information of relevance.

The second file of forensic interest with the application folder is the sync.dat file, which contains a wealth of information relating to the shared folders added by or downloaded to the device under investigation. In particular, the device name could be discerned from the 'device' entry. The 'identity' entry records the identity name ('name') of the device under investigation as well as the private ('private_keys') and public keys ('public_keys') used to establish connections with other devices. A similar finding was observed for the peer identities in the 'identities' entry. We also located a replication of the 'identity' and 'identities' entries in the local-identity-specific */%BitTorrent Sync%/.SyncUser<Random number>/identity.dat* file and the peer-identity-specific */%BitTorrent Sync%/.SyncUser<Random number>/identities/[Certificate fingerprint]* file (with the exception of the private key) respectively. The device or identity name could prove useful in correlating events initiated by a specific identity or device in the log/metadata files or any external data obtained from the peer devices. Meanwhile, the 'access-requests' entry holds a list of metadata pertaining to the identities which sent folder access requests to the device under investigation such as the last used IP addresses in network byte order ('addr'), identity names ('name'), public keys 'public_keys') of the requesting identities, as well as the base32-encoded temporary keys ('invite'), requested folder IDs, requested times ('req_time'), requested permissions ('requested_permissions' where we hypothesised that 2 indicates read only, 3 indicates read and write, and 4 indicates owner), and granted permission ('granted_permissions').

Located within the 'folders' entry of the sync.dat file was metadata relating to the synced folders. It should be noted that this entry will never be empty as it will always contain at least an entry for the identity-specific */%BitTorrent Sync%/SyncUser<Random number>* application folder. Amongst the information of forensic interest recoverable from the 'folders' entry included the folder IDs ('folder_id'), storage paths ('path'), the addition and last modified dates in Unix epoch format, the peer discovery method(s) used to share the synced folders, the access and root certificates keys, whether the folders have been moved to trash, and other information of relevance. We then correlated the folder IDs with the folder IDs located in */%BitTorrent Sync%/SyncUser<Random number>/devices/[Base32-encoded Peer ID]\folders* to determine the shared folders associated with a peer device. To identify the actual (non-encoded) peer ID and corresponding device name, we mapped the encoded peer ID to the pertinent entries in sync.log (see Table 5). Analysis of the access control list ('acl') subentry (of the 'folders' entry) was able to identify the permission information relating to the identities associated with each shared folder, such as the identity names ('name'), public keys ('public_keys'), signature issuers, the times when the identities were linked to a specific shared folder, as well as other information of relevance. We also located the similar details in the folder-specific */%BitTorrent Sync%/.SyncUser<Random number>/folders/[Folder ID]/info.dat* file. The 'peers' subentry (of the 'folders' entry), if available, would provide a practitioner information about the peers associated with the shared folders added by the device under investigation such as the last completed sync time ('last_sync_completed'), last used IP address ('last_addr') in network byte order, device name ('name'), last seen time ('last_seen'), last data sent time ('last_data_sent'), and other relevant information.

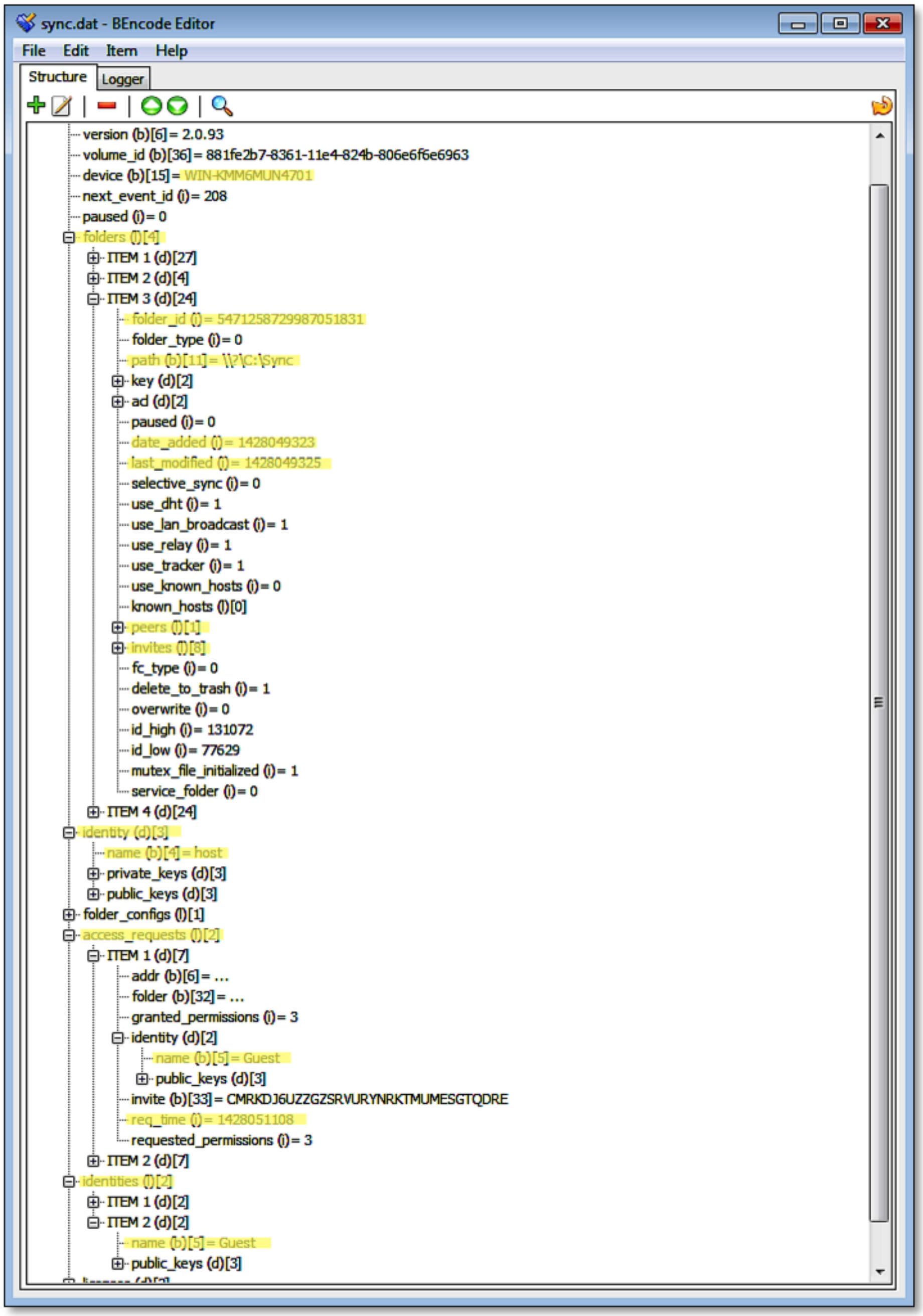

Figure 2: sync.dat.

Another file of interest which can potentially allow a practitioner to recover the sync metadata is the */%BitTorrent Sync%/[share-ID].db* SQLite3 database. This share-ID-specific database describes the content of a share folder (including the */%BitTorrent Sync%/SyncUser<Random number>* application folder) such as the shared filenames or folder names (stored in the 'path' table field of the 'files' table), hashes, and transfer piece registers for the shared files or folders. Once the shared filenames or folder names have been identified, the practitioner could then map the details to the */%BitTorrent Sync%/history.dat* file (which maintains a list of file syncing events appeared in the History windows of the BitTorrent Sync client application) to obtain the sync times in Unix epoch format as well as the associated device names - see Figure 3.

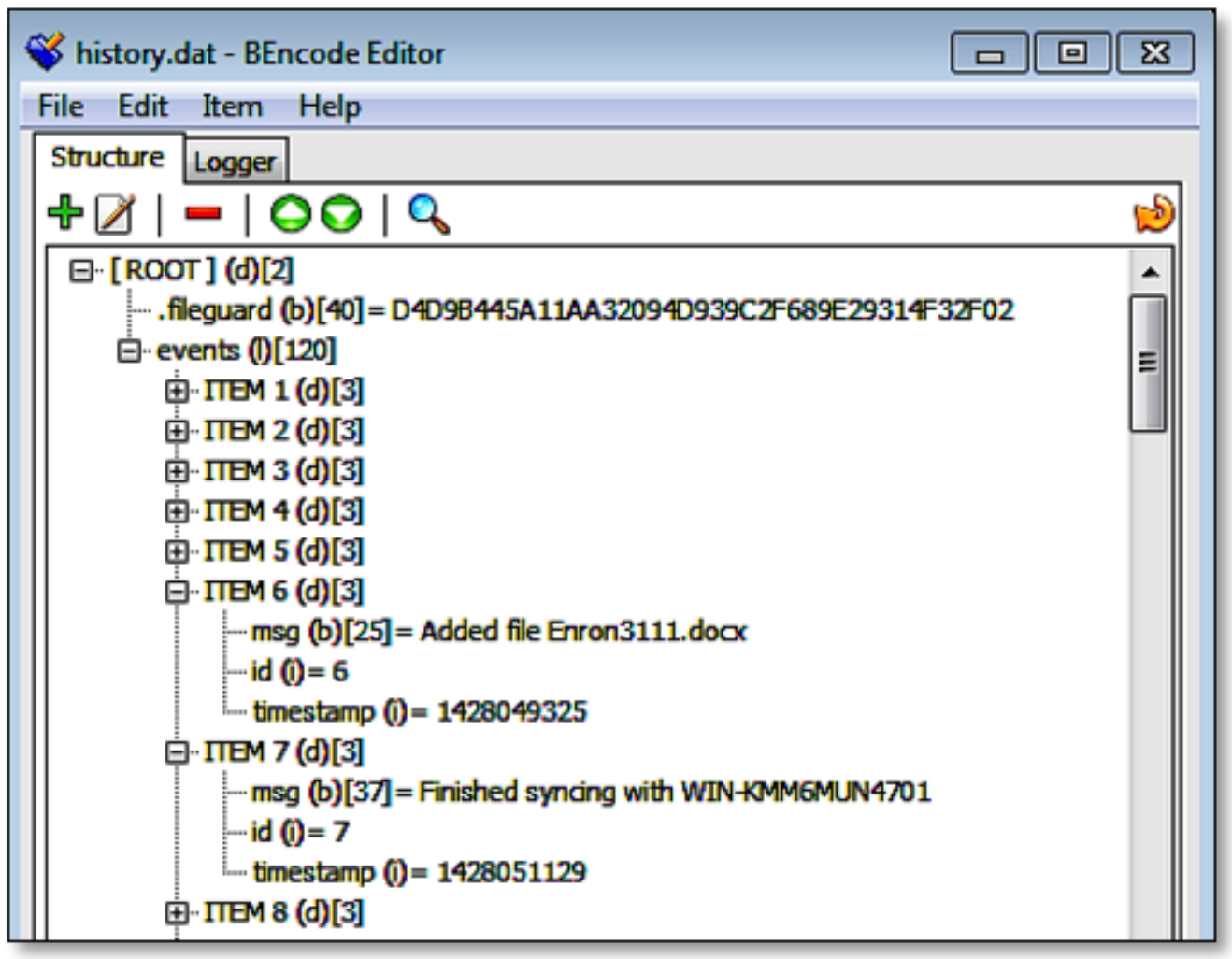

Figure 3: history.dat.

Within the */%BitTorrent Sync%/sync.pid* file there holds the last used process identifier (PID) in plain text. The PID could assist the practitioner in correlating data associated with the BitTorrent Sync client application during RAM analysis (e.g., mapping a string of relevance to the data resided in the memory space of the PID using the 'yarascan' function of Volatility). It is important to note that all the metadata files aforementioned are bencoded (with the exception of the sync.pid file) and the old metadata files will be given an .OLD extension. Moreover, the sync.dat, settings.dat, and history.dat files are protected with a salted file guard key to ensure that only the BitTorrent Sync application can edit the files (Farina et al., 2014).

When BitTorrent Sync was accessed on a Mac OS device, we were able to locate additional references to the client application usage in the preference files located in */Users/[User profile]/Library/Preferences/*. For instance, the com.apple.spotlight.plist file holds the app path and the last used time in plain text (see Figure 4). In the com.bittorrent.Sync.plist file, we recovered supporting information for timeline analysis such as the app version installed, last software update check time, and last started time in Unix epoch format.

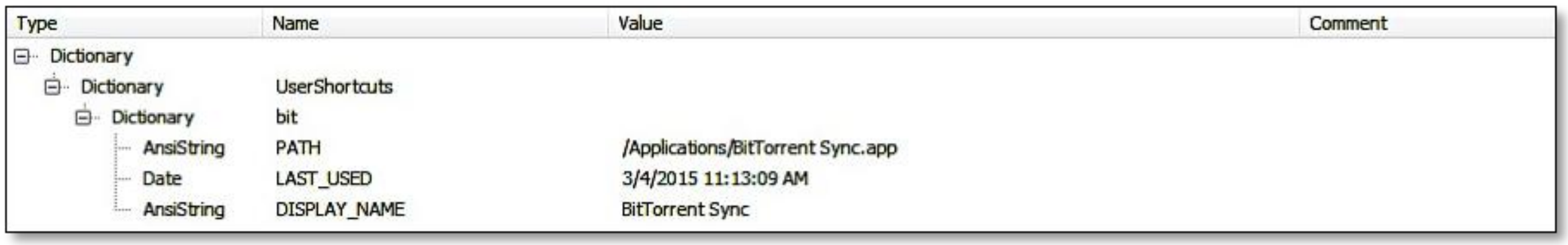

Figure 4: com.apple.spotlight.plist

Disconnecting a shared folder, it was observed that no changes were made to the peer devices, even when the option 'delete files from this device' was selected to permanently delete the sync files/folders from the local device. When we unlinked an identity from the devices investigated, it was identified that the identity-specific */%BitTorrent Sync%/.SyncUser<Random number>* application folder will be deleted from the local device. On the other hand, only the identity-specific metadata will be removed from the 'identity' and 'identities' entries of the local and peer device's settings.dat file respectively.

Undertaking uninstallation of the Windows client application observed discern the synced folders from folders containing the '.sync' subfolder in the directory listing. Our manual uninstallation of the Linux and Mac client applications left no trace of the client application usage/installation in the directory listing, but the deleted files/folders were recoverable from the non-emptied */Users/[User profile]/.Trash* folder of the Mac OSX VM investigated.

Similar to the memory analysis, undertaking data carving of the unallocated space (of the file synchronisation VMs) managed to recover copies of synced files as well as the log and metadata files of forensic interest (e.g., sync.log, sync.dat, history.dat, and settings.dat used by the client applications) intact. A search for the terms 'bittorrent', 'enron', bencode keys specific to the metadata files of relevance, as well as the pertinent log entries was able to locate copies of the recovered files aforementioned in plain text. The remnants remained even after uninstallation of client applications, which suggested that unallocated space is an important source for recovering deleted BitTorrent Sync or synced files.

## 4.2 Log files

Logs play a vital role in reconstructing a criminal scene (Ab Rahman and Choo 2015). BitTorrent Sync stores the logs in the application folder and the filename of which is displayed as 'sync.log'. The default log size is 100MB and can be modified by the user. When the maximum size is reached, the log file is renamed to sync.log.old, and a new sync.log file will be created. As BitTorrent Sync does not implement an encryption algorithm to secure its logs, the logs could be easily accessible using a text editor. The log file is important as it would aid in identifying BitTorrent Sync events around a specific time of incident. Table 5 and Table 6 summarise a list of notable log entries forensic interest from sync.log.

Table 5: Log entries of forensic interest from sync.log.

| Relevance | Examples of log entries obtained in our research |
|---|---|
| Enables a practitioner to identify the BitTorrent Sync version installed on the device under investigation. | • platform: Windows workstation 6.3.0 x86<br>version: 2.0.93 |
| Assist the practitioner in determining the non-encoded peer ID of the device under investigation. | • [2015-04-03 16:18:32] My PeerID: 103B760A3674FE44C4A512B4EF802D452F633F99 |
| A master folder will only be created during identity creation. This potentially allows the practitioner to determine when BitTorrent Sync was first used on a device. | • [2015-04-03 16:19:50] MD[init]: Master Folder: create |
| May assist the practitioner in determining the IP addresses used by the device under investigation. | • [2015-04-03 16:18:30] Using IP address 192.168.220.176<br>• [2015-04-03 16:31:03] Changing IP address from 192.168.220.176 to 192.168.220.143 |
| Informs the practitioner the IP addresses used by the peer devices. | • [2015-04-04 09:05:32] Incoming connection from 192.168.220.176:49734<br>• [2015-04-03 16:51:58] SD[BBAD]: Peer 1: local IP 192.168.220.176:20566<br>• [2015-04-03 16:51:47] SD[BBAD]: Got ping (broadcast: 1) from peer 192.168.220.176:20566 (10DEC8109E524439D9454ABE2BB1475BF7D5A2B5)<br>• Peer 1: 60.50.83.170:49449 10DEC8109E524439D9454ABE2BB1475BF7D5A2B5<br>• [2015-04-05 08:23:56] SF[1F7E] [A2B5]: Found peer 10DEC8109E524439D9454ABE2BB1475BF7D5A2B5 192.168.220.176:49759 direct:1 transport:1 version: 2.0.93 |
| Allows a practitioner to identify the device names of the peer devices. | • [2015-04-05 09:05:32] SF[B5E2] [A2B5]: Got id message from peer **WIN-KMM6MUN4701** (10DEC8109E524439D9454ABE2BB1475BF7D5A2B5) 2.0.93<br>• [2015-04-17 12:51:19] MD[A965]: new device found **WIN-KMM6MUN4701** (CDPMQEE6KJCDTWKFJK7CXMKHLP35LIVV) |
| Since most peer IDs are stored in base32 format in the metadata and configuration files, these log entries would provide a potential method for identification of the actual (non-encoded) peer IDs from the device names. | • [2015-04-05 09:05:32] SF[B5E2] [A2B5]: Got id message from peer WIN-KMM6MUN4701 (**10DEC8109E524439D9454ABE2BB1475BF7D5A2B5**) 2.0.93<br>• [2015-04-15 12:30:31] SD[4F11]: Got ping (broadcast: 1) from peer 192.168.220.146:50523 (**107C1CFB546B565559FE2929E7B7C8804E7302F0**)<br>• [2015-04-17 12:51:19] MD[A965]: new device found WIN-KMM6MUN4701 (**CDPMQEE6KJCDTWKFJK7CXMKHLP35LIVV**)<br>• [2015-04-17 12:51:19] API: callback id=19, value="{ "value": {"peerid":"**CDPMQEE6KJCDTWKFJK7CXMKHLP35LIVV**"}}", can_deferred=0, _delegate=0x1c57d48… |
| May assist the practitioner in determining the share IDs for the shared folders added. | • [2015-04-05 11:37:54] SSLEH[0x15fa28b0]: hello packet { **share:6C25389E651AC160F91ECAF3D9A249C58F6BED15** } has been sent<br>• [2015-04-05 11:37:54] SSLEH[0x08e849e8]: received hello packet, { **share:6C25389E651AC160F91ECAF3D9A249C58F6BED15** }<br>• [2015-04-05 11:47:58] Requesting peers from tracker 52.1.1.135:3000 for share **6C25389E651AC160F91ECAF3D9A249C58F6BED15** |
| Enables identification of the shared folder names/IDs created on the device under investigation. | • [2015-04-04 20:36:45] FC[B5E2]: started periodic scan for **"\\?\C:\Sync"**<br>• [2015-04-05 11:37:57] MD[A965]: [apply] Processing folder **"Sync" (-2775350472753142605)** |
| Assists the practitioner in determining the synced filenames or folder names as well as the addition/creation times. | • [2015-04-05 08:24:17] JOURNAL[22F5]: new torrent created for file Enron3111.txt mt:1418488391 9603FC44BB0F59A822FA3331A1802F880ABA583B<br>[2015-04-05 08:24:17] JOURNAL[22F5]: setting time for file "\\?\C:\Sync\Enron3111.txt" to 1428193457<br>[2015-04-05 08:24:17] JOURNAL[22F5]: insert file "\\?\C:\Sync\Enron3111.txt" = 131072:22982<br>… |
| Informs the practitioner folder names for the deleted folders as well as the deletion times. | • [2015-06-28 23:41:17] Folder being removed from this device and the files at '\\?\C:\Sync' are being removed. |
| Allows the practitioner to determine the local identity's disconnection time. | • [2015-04-05 09:12:01] Master Folder Controller: disconnect master folder |

Table 6: Records of BitTorrent Sync's Application Programming Interface (API) response bodies (in JSON format) of forensic interest from sync.log.

| Relevance | Examples of log entries obtained in our research |
|---|---|
| Provides the practitioner details about the device under investigation such as the peer ID, device name, last online time, last sync completed time, and folder IDs for the shared folders created/added. | • [2015-04-05 09:11:53] API: <-- getmfdevices({ "status": 200, "value": [{ "aod": false, "devicename": "WIN-KMM6MUN4701", "folders": [ { "added": true, "id": -7338009380596345790, "mode": 1 }, { "added": true, "id": 3964779361527927184, "mode": 1 }, { "added": true, "id": 4780923171276619705, "mode": 1 }, { "added": true, "id": 5471258729987051831, "mode": 1 } ], "id": "CDPMQEE6KJCDTWKFJK7CXMKHLP35LIVV", "lastseen": 1428196287, "lastsynccompleted": 1428196287, "name": "WIN-KMM6MUN4701", "online": true, "self": false, "syncerr": 0, "syncerrmsg": "", "userid": "" } ] })… |
| Assists the practitioner in determining the pending user requests sent to the device under investigation including the folder IDs (if any), the times when the requests were sent, access permissions, as well as the requester's IP addresses and certificate fingerprints. | • [2015-04-03 16:51:48] API: <-- getpendingrequests({ "status": 200, "value": [ { "access_level": 3, "id": "5471258729987051831", "ip": "192.168.220.176", "license": false, "readwrite": true, "time": 1428051108, "user_identity": { "devicename": "device", "fingerprint": "2UMI566O3XAE7BB2V3N3YWWECJ3TCGJHMRGZTVLN2SZY276QI4AQ", "username": "Guest" } } ] })… |
| May assist a practitioner in determining the folder names, folder IDs, storage paths, folder sizes, timestamp information, as well as peer device names, peer IDs, and fingerprints associated with the shared folders added by or downloaded to the device under investigation. | • [2015-04-05 09:05:37] API: <-- getsyncfolders({ "folders": [ { "access": 4, "archive": "C:\\Sync\\.sync\\Archive", "archive_files": 3, "archive_size": 153187, "date_added": 1428049323, "down_eta": 0, "down_speed": 0, "down_status": 100, "error": 0, "files": 3, "folderid": "5471258729987051831", "has_key": true, "indexing": false, "ismanaged": true, "iswritable": true, "last_modified": 1428053450, "name": "Sync", "path": "C:\\Sync", "paused": false, "peers": [ { "direct": true, "downdiff": 0, "id": "10DEC8109E524439D9454ABE2BB1475BF7D5A2B5", "isonline": true, "lastreceivedtime": 0, "lastsenttime": 1428051120, "lastsynctime": 1428051129, "name": "WIN-KMM6MUN4701", "updiff": 0, "userid": "UQO52P4G5O2QU6OOGX3AS7R6RUAU22JBBWJ4H2CYNXHRO3KIRVBQ" }], "size": 321638, "status": "314.0 kB in 3 files", "stopped": false, "synclevel": 2, "up_eta": 0, "up_speed": 0, "up_status": 100, "users": [{ "access": 3, "id": "2UMI566O3XAE7BB2V3N3YWWECJ3TCGJHMRGZTVLN2SZY276QI4AQ", "name": "Guest" } ] },<br>… |
| Informs the practitioner the storage path for the device under investigation. | • [2015-04-03 16:43:13] API: <-- getfoldersstoragepath({ "status": 200, "value": "C:\\Users\\anonymous\\BitTorrent Sync" })<br>• [2015-04-05 09:05:33] API: <-- setfoldersstoragepath({ "path": "C:\\Users\\anonymous\\BitTorrent Sync", "status": 200 }) |
| Allows the practitioner to identify the folder name, path, and timestamp references for the shared folders added by the device under investigation. | • [2015-04-04 20:27:22] API: --> addsyncfolder(path=C%3A%5CSync&selectivesync=false&t=1428150442927) |
| Contains copy of history.dat file (see section 4.1) at the time of request. | • [2015-04-05 08:33:06] API: <-- history({ "status": 200, "value": [{ "id": 39, "msg": "WIN-KMM6MUN4701 updated file Enron3111.zip", "time": 1428193777 }, { "id": 38, "msg": "WIN-KMM6MUN4701 updated file Enron3111.txt", "time": 1428193777 }, { "id": 37, "msg": "Remote peer removed file Enron3111.rtf", "time": 1428193777 }, { "id": 13, "msg": "Added file Enron3111.docx", "time": 1428153859 }… |

### 4.3 Physical memory

Memory analysis is invaluable for recovering information which would otherwise be lost (Canlar et al., 2013). Memory analysis in this research included undertaking data carving using Photorec, keyword search using a Hex editor, and contextualising the RAM contents using Volatility. During RAM analysis, a practitioner must aware that memory changes frequently according to users' activities and will be wiped off as soon as the system is shut down. Thus, the data remnants identified this research does not represent those recoverable in a typical "real world" circumstance but serves as a guideline for detecting possible recoverable evidences.

Analysis of the running processes using the ‘pslist’ function of Volatility was able to recover the process name associated with the BitTorrent Sync client application (e.g., ‘BitTorrent Sync.exe’ for Windows OS, ‘BitTorrent Sync’ for Linux OS, and ‘BitTorrent Sync’ for Mac OS), which included the process identifier (PID), parent process identifiers (PPID) as well as the process initiation time; Examinations of the network details using the ‘netscan’ or ‘netstat’ function of Volatility determined that the network and socket information such as the transportation protocols used, local and remote IP addresses (including the IP addresses of the peer discovery methods used and the peer nodes), socket states, as well as the timestamp information could be recovered from the RAM (see Figure 5).

| Offset(P) | Proto | Local Address | Foreign Address | State | Pid | Owner | Created |
|---|---|---|---|---|---|---|---|
| 0x7b8017e0 | UDPv4 | 127.0.0.1:16448 | *:* | | 956 | BTSync.exe | 2015-04-03 08:18:31 UTC+0000 |
| 0x7bb71e60 | UDPv4 | 0.0.0.0:0 | *:* | | 956 | BTSync.exe | 2015-04-03 12:29:56 UTC+0000 |
| 0x7b54f790 | TCPv4 | 192.168.220.143:49765 | 60.50.83.170:51316 | CLOSED | 956 | BTSync.exe | |
| 0x7b564320 | TCPv4 | 192.168.220.143:49766 | 67.215.229.106:3000 | ESTABLISHED | 956 | BTSync.exe | |
| … | | | | | | | |
| 0x7c0edec0 | UDPv4 | 0.0.0.0:0 | *:* | | 956 | BTSync.exe | 2015-04-03 12:29:56 UTC+0000 |
| … | | | | | | | |
| 0x7c28cc00 | TCPv4 | 192.168.220.143:49740 | 52.1.1.135:3000 | CLOSED | 956 | BTSync.exe | |
| … | | | | | | | |
| 0x7d6d2730 | UDPv4 | 0.0.0.0:0 | *:* | | 956 | BTSync.exe | 2015-04-03 12:29:56 UTC+0000 |
| … | | | | | | | |
| 0x7d850290 | TCPv4 | 0.0.0.0:39259 | 0.0.0.0:0 | LISTENING | 956 | BTSync.exe | |
| 0x7d11b5f0 | TCPv4 | 192.168.220.143:49742 | 52.1.40.103:3000 | CLOSED | 956 | BTSync.exe | |
| 0x7d604c30 | TCPv4 | 192.168.220.143:49753 | 10.63.132.243:3001 | CLOSED | 956 | BTSync.exe | |
| … | | | | | | | |
| 0x7dd47cb0 | UDPv4 | 0.0.0.0:0 | *:* | | 956 | BTSync.exe | 2015-04-03 08:18:31 UTC+0000 |
| 0x7da277c0 | TCPv4 | 192.168.220.143:49741 | 52.0.104.40:3000 | ESTABLISHED | 956 | BTSync.exe | |
| 0x7da99590 | TCPv4 | 192.168.220.143:39259 | 192.168.220.176:49582 | CLOSED | 956 | BTSync.exe | |

Figure 5: An excerpt of BitTorrent Sync network information recovered using the ‘netscan’ function of Volatility.

Undertaking data carving of the RAM captures and swap files determined that only the images used by the client application and synced files could be recovered. However, a search for the term ‘btsync’ or ‘bittorrent sync’ was able to recover the complete text of the log and metadata files of forensic interest (e.g., sync.log, sync.dat, history.dat, and settings.dat) in the RAM in plain text. In cases when the original file has been deleted, a Yarascan search for the text from the remnants could help attribute the remnants to the BitTorrent Sync or other processes of relevance to identify its origin. Figure 6 illustrates an occurrence of history.dat in the memory space of ‘BitTorrent Sync.exe’ of the Windows 8.1 VM investigated. The term ‘bittorrent’, bencode keys specific to the metadata files of relevance (see Section 4.1 and Section 4.2), and log entries as identified in Table 5 could be used to narrow down the search space.

```
Rule: r1
Owner: Process BTSync.exe Pid 956
0x008a0e7b  33 31 31 31 2e 74 78 74 22 2c 20 22 74 69 6d 65   3111.txt",."time
0x008a0e8b  22 3a 20 31 34 32 38 30 35 32 33 35 31 20 7d 2c   ":.1428052351.},
0x008a0e9b  20 7b 20 22 69 64 22 3a 20 31 38 2c 20 22 6d 73   .{."id":.18,."ms
0x008a0eab  67 22 3a 20 22 57 49 4e 2d 4b 4d 4d 36 4d 55 4e   g":."WIN-KMM6MUN
0x008a0ebb  34 37 30 31 20 61 64 64 65 64 20 66 69 6c 65 20   4701.added.file.
0x008a0ecb  45 6e 72 6f 6e 33 31 31 31 2e 72 74 66 22 2c 20   Enron3111.rtf",.
0x008a0edb  22 74 69 6d 65 22 3a 20 31 34 32 38 30 35 32 33   "time":.14280523
0x008a0eeb  35 31 20 7d 2c 20 7b 20 22 69 64 22 3a 20 31 37   51.},.{."id":.17
0x008a0efb  2c 20 22 6d 73 67 22 3a 20 22 57 49 4e 2d 4b 4d   ,."msg":."WIN-KM
0x008a0f0b  4d 36 4d 55 4e 34 37 30 31 20 61 64 64 65 64 20   M6MUN4701.added.
0x008a0f1b  66 69 6c 65 20 45 6e 72 6f 6e 33 31 31 31 2e 70   file.Enron3111.p
0x008a0f2b  64 66 22 2c 20 22 74 69 6d 65 22 3a 20 31 34 32   df",."time":.142
0x008a0f3b  38 30 35 32 33 35 31 20 7d 2c 20 7b 20 22 69 64   8052351.},.{."id
0x008a0f4b  22 3a 20 31 36 2c 20 22 6d 73 67 22 3a 20 22 57   ":.16,."msg":."W
0x008a0f5b  49 4e 2d 4b 4d 4d 36 4d 55 4e 34 37 30 31 20 61   IN-KMM6MUN4701.a
0x008a0f6b  64 64 65 64 20 66 69 6c 65 20 45 6e 72 6f 6e 33   dded.file.Enron3
```

Figure 6: Copy of history.dat file recovered from the memory space of ‘BitTorrent Sync.exe’.

Unsurprisingly, a keyword search for Enron-related keywords (e.g., ‘Enron’) was able to locate the complete text of the sample files in the RAM captures of all the file synchronisation VMs investigated. Although the login credentials were encrypted, we were able to recover the username (login email) and password for the Linux client application’s web GUI following the strings ‘username=’ and ‘nwpwd=’ in the RAM respectively. These appeared to be remnants from the form input field of the Linux client application’s web GUI; an example is shown in Figure 7. In addition, we also located several password hits in the similar fragments containing the login email in the memory space of BitTorrent Sync.  In a real life scenario, this could potentially provide the practitioner the opportunity to extrapolate the login password from the non-dictionary or alphanumeric terms surrounding the email string in the memory space of BitTorrent Sync.

```
2DC1DA20  00 00 00 8F 48 54 54 50 3A 68 74 74 70 3A 2F 2F  ....HTTP:http://
2DC1DA30  31 32 37 2E 30 2E 30 2E 31 3A 38 38 38 38 2F 67  127.0.0.1:8888/g
2DC1DA40  75 69 2F 3F 74 6F 6B 65 6E 3D 74 49 36 44 52 51  ui/?token=tI6DRQ
2DC1DA50  46 59 43 64 63 6F 2D 4E 73 54 65 74 62 56 34 31  FYCdco-NsTetbV41
2DC1DA60  4B 73 6D 79 6D 79 78 55 64 6E 73 44 7A 77 31 6A  KsmymyxUdnsDzw1j
2DC1DA70  46 4C 6F 55 72 62 79 68 48 51 6A 76 66 63 4F 79  FLoUrbyhHQjvfcOy
2DC1DA80  70 6B 48 6C 55 41 41 41 41 41 26 61 63 74 69 6F  pkHlUAAAAA&actio
2DC1DA90  6E 3D 73 65 74 6C 69 63 65 6E 73 65 61 67 72 65  n=setlicenseagre
2DC1DAA0  65 64 26 76 61 6C 75 65 3D 74 72 75 65 26 75 73  ed&value=true&us
2DC1DAB0  65 72 6E 61 6D 65 3D [redacted]  ername=[redacted]
2DC1DAC0  [redacted] 26 6E 65  [redacted]&ne
2DC1DAD0  77 70 77 64 [redacted]  wpwd=[redacted]
2DC1DAE0  [redacted] 64 70 77 64 3D 26 74 3D 31 34 32  [redacted]pwd=&t=142
2DC1DAF0  38 30 35 35 31 36 36 39 39 37 00 72 65 71 75 65  8055166997.reque
2DC1DB00  73 74 2D 6D 65 74 68 6F 64 00 47 45 54 00 72 65  st-method.GET.re
2DC1DB10  73 70 6F 6E 73 65 2D 68 65 61 64 00 48 54 54 50  sponse-head.HTTP
2DC1DB20  2F 31 2E 31 20 32 30 30 20 4F 4B 0D 0A 43 6F 6E  /1.1 200 OK..Con
2DC1DB30  74 65 6E 74 2D 4C 65 6E 67 74 68 3A 20 34 32 0D  tent-Length: 42.
2DC1DB40  0A 43 6F 6E 74 65 6E 74 2D 54 79 70 65 3A 20 61  .Content-Type: a
2DC1DB50  70 70 6C 69 63 61 74 69 6F 6E 2F 6A 73 6F 6E 3B  pplication/json;
2DC1DB60  20 63 68 61 72 73 65 74 3D 75 74 66 2D 38 0D 0A   charset=utf-8..
2DC1DB70  43 61 63 68 65 2D 43 6F 6E 74 72 6F 6C 3A 20 6E  Cache-Control: n
2DC1DB80  6F 2D 63 61 63 68 65 0D 0A 00 00 00 00 00 00 00  o-cache.........
```

Figure 7: Username and password recovered from the RAM of Ubuntu OS.

## 4.4 Thumbnail cache

Analysis of the Windows thumbcache (stored under *%AppData%\Local\Microsoft\Windows\Explorer*) recovered copies of thumbnail images for the BitTorrent Sync client application and its download site (e.g., BitTorrent Sync logo and image icons), indicative of BitTorrent Sync usage. Examinations of the thumbnail cache from the file synchronisation VMs only revealed copies of thumbnail images for the synced files from the Windows 8.1 and Mac OS VMs. We could discern the thumbnail cache from the 'folder' table field (of the 'files' table) which made reference to 'BitTorrent Sync' in the */private/var/folders[Random subfolder]*/*[Random subfolder]/C/com.apple.QuickLook.thumbcache/index.sqlite* database of the Mac OS VM investigated. The timestamp references recorded alongside the thumbnail cache would assist a practitioner to identify the last accessed or deletion (only in a Mac OS device; see Figure 8) date of a sync file or folder. Moreover, the presence of the thumbnail cache in the file deletion and uninstallation VMs suggests that thumbnail cache is an important source evidence to recover deleted images; this seems to agree with the findings of Quick et al. (2014b). Analysis of the Ubuntu VMs did not recover any thumbnail instance relevant to BitTorrent Sync.

| | folder | file_name | fs_id | version |
|---|---|---|---|---|
| | Filter | Filter | Filter | Filter |
| 1 | /Users/alice/Downloads | About Downloads.lpdf | /.file/id=6571367.404398/ | (BLOB) |
| 2 | /Users/alice/BitTorrent Sync/Sync | Enron3111.pdf | /.file/id=6571367.409486 | (BLOB) |
| 3 | /Users/alice/BitTorrent Sync/Sync | Enron3111.jpg | /.file/id=6571367.409485 | (BLOB) |
| 4 | /Users/alice/.Trash | Enron3111 12.07.57 AM.pdf | /.file/id=6571367.409694 | (BLOB) |
| 5 | /Users/alice/.Trash | Enron3111 12.07.57 AM.jpg | /.file/id=6571367.409693 | (BLOB) |
| 6 | /Users/alice/.Trash | Enron3111 12.07.57 AM.rtf | /.file/id=6571367.409695 | (BLOB) |
| 7 | /Users/alice/.Trash | Enron3111 12.07.57 AM.docx | /.file/id=6571367.409690 | (BLOB) |
| 8 | /Users/alice/.Trash | Enron3111 12.08.03 AM.txt | /.file/id=6571367.409702 | (BLOB) |

Figure 8: Thumbnail information recovered from the index.sqlite database of Mac OS' thumbcache folder.

## 4.5 Windows Registry

Windows registry provides a rich source of information associated with installed programs (Do et al. 2014). Although five hives could be seen in the registry, only HKEY_USERS (HKU) and HKEY_LOCAL_MACHINE (HKLM) hives are tangibly real since the remaining are merely symbolic links to the two master keys (Farmer, 2007). Our analysis of the HKLM hive determined that the BitTorrent Sync installation could be detected from the presence of the *HKLM\SOFTWARE\BitTorrent\Sync* key, and the installation path could be discerned from the ‘SyncPath’ subkey. In addition, the *HKLM\SOFTWARE\Microsoft\Windows\CurrentVersion\Uninstall\BitTorrent Sync* key could provide supporting information for the installation such as the display icon’s path, display name, BitTorrent Sync version installed, installation and uninstaller paths, and other entries of relevance.

Similar to any other Windows application, when the BitTorrent Sync client application was started up, we located full path reference for the BitTorrent Sync executable file in *HKU\<SID>\Software\Classes\Local Settings\Software\Microsoft\Windows\Shell\MuiCache*, indicative of recent BitTorrent Sync usage. Further evidence to indicate the client application usage could be ascertained from the occurrence of ‘BitTorrent Sync: ""%Program Files%\BitTorrent Sync\BitTorrent Sync.exe" /MINIMIZED"’ entry alongside the last executed time in *Software\Microsoft\Windows\CurrentVersion\Run*.

Another registry key of forensic interest is the *Software\Microsoft\Windows\CurrentVersion\Explorer\ComDig32*, which keeps track of a list of filename references (e.g., filenames for the executable and synced files) associated with the BitTorrent Sync client application as well as the timestamp information during the last usage. According to Carvey (2014), the ‘CIDSizeMRU’ (MRU is the abbreviation for Most-Recently-Used) subkey maintains a list of recently used applications, the ‘OpenSaveMRU’ registry subkey records list of files that have been opened or saved within a Windows shell dialog box, and the ‘LastVisitedMRU’ subkey is responsible for tracking specific executable files used by an application to open the files documented in the ‘OpenSaveMRU’ subkey. Other evidence indicating the BitTorrent Sync client application usage includes the presence of entries referencing the link file as well as the last executed time in *Software\Microsoft\Windows\CurrentVersion\Explorer\UserAssist.*

### 4.6 Prefetch files

Examination of the prefetch files located two prefetch files for BitTorrent Sync, namely ‘BITTORRENT_SYNC.EXE.pf’ and ‘BITTORRENT SYNC.exe.pf’. Amongst the information of forensic interest recoverable from these files include the executable path, a number of times the application has been loaded, as well as the last run time which are useful to supplement timeline analysis. However, no prefetch instance was located for the synced files in our experiments. The presence of the prefetch files in the uninstall VMs (H1.1.1.3 and H1.1.1.3) implies that there will be BitTorrent Sync references remaining in the prefetch files to indicate its use after uninstallation of the client application.

### 4.7 Link files

Link (.lnk) files are shortcut metadata files used by Windows to maintain a list of linked paths relating to a file (commonly the paths where the original files are located), associated timestamps (created, written, and last accessed times), and file sizes (original and modified) which are useful to identify the origin of a file (Microsoft, 2015). An inspection of the directory listings located instances of link file for *%Program Files (x86)%\BitTorrent Sync\BitTorrent Sync.exe* at *%Users%\Public\Desktop\BitTorrent Sync.lnk* and *%Program Data%\Microsoft\Windows\Start Menu\BitTorrent Sync.lnk*, and its presence may be indicative BitTorrent Sync installation.

## 5. Sync analysis on mobile clients

With the growing use of mobile handheld devices, mobile client artefacts can prove an invaluable evidence source in digital forensics investigations (Dezfouli et al., 2015). In this section, we present the BitTorrent Sync findings on iPhone 4 running iOS 7.1.2 and a HTC One X running Android KitKat 4.4.4.

### 5.1 BitTorrent Sync analysis on iOS 7.1.2

Analysis of the directory listing revealed that the iOS app installation could be discerned from the presence of the */private/var/mobile/Applications/[Unique SHA-1 identifier for the BitTorrent Sync iOS app]/BitTorrent Sync.app* file. The application and storage folders (similarly to those presented for the computer applications as highlighted in Section 4.1) could be located at */private/var/mobile/Applications/[Unique SHA-1 identifier for the BitTorrent Sync iOS app]/Documents/BitTorrent Sync* and */private/var/mobile/Applications/[Unique SHA-1* *identifier for the BitTorrent Sync iOS app]/Documents/Storage* respectively. Notice that the sync files/folders will only be downloaded when viewed in the app. Within the */private/var/mobile/Applications/[Unique SHA-1 identifier for the BitTorrent Sync iOS app]/iTunesMetadata.plist* file there maintains a list of mobile-specific metadata associated with the Symform app such as the Apple ID used to purchase the app, the purchase date, the BitTorrent Sync version installed. Alternatively, copies of the application and storage folders aforementioned could be located in */User/Applications/[Unique SHA-1 identifier for the BitTorrent Sync iOS app]/.*

Inspecting the log files, it was determined that the app installation could be distinguished from entries referencing 'BitTorrent Sync' in the */private/var/mobile/Library/Logs/MobileInstallation/mobile_installation.log.#*, which includes the installation time. Meanwhile, the */private/var/mobile/Library/Logs/Powerlog.powerlog* maintains a list of power and network consumption details associated with the BitTorrent Sync app; useful when seeking to determine the app usage pattern. An example of the log entry is as follows:

*"05/05/15 01:50:46 [Network Connections Symptoms] procName=BitTorrent Sync; bundleName=com.bittorent.BitTorrent-Sync; wifi-in=5461675bytes; wifi-out=1206024bytes; cell-in=435717bytes; cell-out=626892bytes; sinceTime=03/19/15 01:08:39"*

Undertaking uninstallation of the iOS app determined that all the files and folders of forensic interest aforementioned were removed, with the exception of the BitTorrent Sync entries located within the OS log files of relevance.

## 5.2 BitTorrent Sync analysis on Android KitKat 4.4

The installation of the Android BitTorrent Sync app resulted in the creation of */data/data/com.bittorrent.sync* app folder. The file of particular interest with this folder is the */data/data/com.bittorrent.sync/shared_prefs/preferences.xml*, which keeps track of a list of paths for the shared folder added by or downloaded to the device under investigation in the 'folders' entry, a count of the number of times the app was run in the 'number_of_runs' entry, and other information of relevance. Another file of particular interest is the */data/data/com.bittorrent.sync/SyncStatistics.xml*. Within this file there holds a list of sync statistics associated with the app such as the last sync time (which could be discerned from the 'last_sent_time' entry).

The application and storage folders could be located at */data/data/com.bittorrent.sync/files/.sync* and */Home/Download/BitTorrent Sync* respectively. Inspecting the directory listing, it was observed that these folders share the same directory structure as those presented for the computer applications (see Section 4.1), and the shared folders could be easily distinguished from the '.sync' subfolder. Similar to the iOS app, the sync files/folders were only downloaded (to the storage folder) when viewed. Only the storage as well as locally added shared folders remained after uninstallation of the app.

# 6. Network analysis

Analysis of the network captures determined that the network traffics were encrypted (due AES-128 encryption from the client) and the synced files were not recovered. However, we were able to recover the peer discovery packets that contain information of forensic interest such the IP addresses, port numbers, peer IDs, and share IDs associated with the host and the peers.

Unless configured otherwise, BitTorrent Sync uses tracker servers for peer discovery by default. It was determined that a client utilising this peer discovery method will first send a tracker request packet to the tracker servers, one each for disparate share IDs, to retrieve the peer list. The tracker request packet could be discerned from the term 'get_peers' and amongst the information of forensic interest recoverable from the packet includes the local IP address, port number, peer ID, as well as the 20 or 32 bytes share ID for the requesting folder (see Table 7). Note that the act of requesting a peer lookup also serves to register the requesting client as a source (Scanlon et al., 2014). Upon receiving the tracker request packet from the requesting client, the tracker server will respond the requesting client with a tracker response packet containing a peer list which shares the same share ID (see Table 8). The tracker response packet will always contain at least the originating peer information and, hence, will never be empty.

Table 7: Details of tracker request packet of BitTorrent Sync (Information of forensic interest is bolded).

| Bencode key and value pairs | Description |
|---|---|
| µTP header type 0 | µTP header type 0 which signifies regular data packet (BitTorrent Inc., 2009) |
| d | Start of bencode dictionary key:value pairs |
| 2:la | Label identifier for local address |
| **6:[local IP:Port number]** | **Local IP address (4 bytes) and port number (2 bytes) combination in network byte order** |
| 2:lp | Label identifier for local port number |
| **i[5-byte port number]e** | **5 bytes local port in integer format** |
| 1:m | Label identifier for message |
| 9:get_peers | Label identifier for 'get_peers' message type |
| 4:peer | Label identifier for local peer |
| **20:[20-byte peer ID]** | **20 bytes local peer ID** |
| 5:share | Label identifier for share ID |
| **20:[20-byte Share ID]e OR 32: [32 bytes share ID]** | **20 or 32 bytes share ID** |

Table 8: Details of tracker response packet of BitTorrent Sync (Information of forensic interest is bolded).

| Bencode key and value pairs | Description |
|---|---|
| µTP header type 0 | µTP header type 0 which signifies regular data packet. |
| 00 | Null |
| d | Start of bencode dictionary key:value pairs |
| 2:ea | Label identifier for external address of the requesting client. |
| **6:[external IP:Port number of the requesting client]** | **External IP address (4 bytes) and port number (2 bytes) combination of the requesting client in network byte order** |
| 1:m: | Message label identifier |
| 5:peers | Label identifier for 'peers' message type |
| 5:peers | Label identifier for 'peers' message type |
| l | Start of peer list |
| d | Start of bencode dictionary key:value pairs for the peer list |
| 1:a | Label identifier for external address |
| **6:[external IP:Port number]** | **External IP address (4 bytes) and port number (2 bytes) combination of the corresponding peer in network byte order** |
| 2:la | Label identifier for internal address |
| **6:[internal/local IP:Port]** | **Internal IP address (4 bytes) and port number (2 bytes) combination of the corresponding peer in network byte order** |
| 1:p | Label identifier for peer ID |
| **20:[20-byte peer ID]e** | **Peer ID of the corresponding peer** |
| e | End of peer list |
| 5:share | Label identifier for share ID |
| **20:[20-byte share ID]e OR 32: [32 bytes share ID]** | **20 or 32 bytes share ID** |
| **4:time:i[timestamp]e** | **4 bytes timestamp in integer and format (Unix epoch)** |

When the local peer discovery method (Search LAN) was used, it was observed that the requesting client first broadcasted the multicast packets to the IP address 239.192.0.0:3838 for revealing its presence. The information of forensic interest recovered from the packets include the local port number, peer ID, and the the 20 or 32 bytes share IDs for the requesting folders (see Table 9). Whilst separate multicast packet was used for disparate share IDs in the older BitTorrent Sync version (Scanlon et al., 2014a, 2014b, 2015), it was identified that the share IDs were broadcasted as a list using a same multicast packet in our research. The peers receiving the multicast message containing the share ID(s) they possess then responded the requesting client with a multicast response packet which shares the same format as the multicast packet.

Table 9: Details of multicast ping packet of BitTorrent Sync (Information of forensic interest is bolded).

| Bencode key and value pairs | Description |
|---|---|
| BSYNC | BSYNC header |
| 00 | Null |
| D | Start of bencode dictionary key:value pairs |
| 1:m | Message label identifier |
| 4:ping | Message type 'ping' |
| 4:peer | Label identifier for peer |
| 4:port | Label identifier for local port number |
| **I[Port number]e** | **Local port number in integer format** |
| **20:[20-byte peer ID]** | **20 bytes local peer ID** |
| 6:shares | Label identifier for share ID |
| L | Start of share ID list |
| **20:[20-byte peer ID]e OR**<br>**32: [32 bytes share ID]e** | **20 or 32 bytes share ID** |
| E | End of share ID list |

Other than inspecting the packet formats, the peer discovery methods could be discerned from the IP address and port number combinations maintained in the sync.conf file, which is accessible through *http://config.usyncapp.com/sync.conf*. Table 10 summarises the latest IP addresses and port numbers obtained from the sync.conf file downloaded in our research. Alternatively, we could also locate copies of the sync.conf file in the sync.log file as outlined in Table 5. Since our experiments only limited to the default peer discovery options, only the tracker server and local peer discovery methods were used. Comprehensive analysis of the network protocols used by the remaining peer discovery methods is beyond the scope of this paper.

Table 10: IP addresses and port numbers used by BitTorrent Sync.

| Sync preferences | IP address(es) | Port number | URL |
|---|---|---|---|
| Use tracker server | 52.0.104.4.40, 52.0.102.230, 52.1.40.103, 52.1.1.135 | TCP and UDP port 3000 | t.usyncapp.com |
| Use relay server when required | 67.215.231.242, 67.215.229.106 | TCP and UDP port 3000 | r.usyncapp.com |
| Search LAN | 239.192.0.0 | UDP port 3838 | |
| Automatic ports mapping over UPnP and NAT-PMP UDP multicast | 239.255.255.250 | port 1900 UDP unicast to default gateway port 5351 | |
| mobile_push_proxies | 54.235.182.157 | TCP and UDP port 3000 | |

## 7. BitTorrent Sync investigative methodology

Based on our investigations of BitTorrent Sync, we outline a process that can be used to guide future forensic practitioners. In order to minimise the risk of the evidence from being questioned in court of law, it is imperative that the process adhere to generally accepted forensic principles, standards, guidelines, procedures and best practices (NIJ, 2004; Kent et al., p.5, 2006; ACPO, 2011). Mckemmish (1999), for example, defines four rules for digital forensics, namely minimal handing of the original, account for any changes, comply with the rules of evidence, and not to exceed knowledge. Similarly, the digital forensic principles of the United Kingdom Association of Chief Police Officers (ACPO) specify that: no action should change data, when it is necessary to access original data the persons accessing data should be competent to do so, a record of processes should be made, and the investigator in charge is responsible to ensure the principles are adhered to. Meanwhile, the NIST requires a digital forensics framework to support collection, examination, analysis, and reporting of (digital) evidences (Kent et al., 2006).

A cloud forensics framework that could provide a sound basis for compliance with the above principles is the cloud forensics framework of Martini and Choo (2012), since the framework is built on the digital forensics frameworks of McKemmish (1999) and NIST (Kent et al., 2006). Hence, similar to the approach of Quick et al. (2014b), we subsequently mapped the BitTorrent Sync investigation process to the framework of Martini and Choo (2012) to provide an operational methodology for forensic practitioners. This involves the following operation methodology:

### Identification and preservation

In the first stage, a practitioner identifies and seizes physical sources of evidence relevant to BitTorrent Sync investigation such as desktop computers, laptops, and mobile devices. Before attempting to preserve the data, it is encouraged to isolate the system from network access to avoid further human intervention. Then, a proper and timely preservation of the hard drive and volatile data captures using standard forensic image formats (E01, bit-for-bit dd image, etc.) should be undertaken, as soon as practical. If physical acquisition is not possible, continue with the next (collection) stage.

**Collection**

Before collecting the data of relevance, the practitioner should determine whether BitTorrent Sync is installed on the device under investigation. This could be accomplished by determining the presence of the BitTorrent Sync installation directory and the pertinent entries in the OS log as well as other OS-specific files e.g., the registry, link, and prefetch files of Windows OS. Then, collection commences with extracting (exporting) the BitTorrent Sync installation, application, and sync/storage (which could be discerned from the '.sync' subfolder) folders from the forensic image using a forensic browser. In addition, the practitioner is encouraged to carve and collect the files recorded within the unallocated space, RAM dump, and swap file using a data recovery software for supporting information. If physical acquisition is not undertaken in the previous (identification and collection) stage, logical acquisition should be undertaken of the files of relevance (on the device under investigation) using a live acquisition software and properly preserved in a logical container such as L01, AD1, or CTR format. The collection should also include OS-specific and web browser files of forensic interest such as $MFT, $LogFile, $UsnJrnl, prefetch files, thumbnail cache, link files, log files, as well as other files as highlighted in Section 4.

**Examination and analysis**

In this stage, the practitioner examines the collected files. Both indexed and non-indexed as well as Unicode and non-Unicode string search should be included as part of the keyword search. The analysis involves the following steps:

1. Analyse the settings.dat and sync.dat files to determine the BitTorrent Sync installation time, peer ID, and device name of the device under investigation.
2. Next, analyse the 'identity' and 'identities' entries of the sync.dat file to determine the identity names of the device under investigation and the associated peer devices.
3. Analyse the 'access_request' entry of the sync.dat file to obtain the identities which sent folder access requests to the device under investigation as well as the IP addresses, folder IDs, request times, and request permission types.
4. Analyse the 'folder' entry of the sync.dat file to determine the folder IDs, folder names, last added and modified times, as well as the paths for the shared folders added by or downloaded to the device under investigation.
5. Map the information obtained in step 4 to the physical copies of the shared folders to determine the share ID (from the 'ID' file stored within the '.sync' subfolder), sync filenames or folder names (including the filenames/folder names for the deleted files/folders resided in the archive folder), as well as the timestamp information associated with each shared folder.
6. Match the share ID(s) obtained in step 5 with the [share-ID].db database in the application folder to determine the filenames or folder names associated with each share.
7. Match the synced filenames or folder names determined in step 5 with the history.dat file (or the history.dat entries recorded in sync.log) to determine the device names associated with a particular synced filename or folder name.
8. Correlate the folder IDs obtained in step 4 with the folder IDs located in */%BitTorrent Sync%/SyncUser<Random number>/devices/[Base32-encoded Peer ID]/folders/* to determine the shared folders associated with a specific peer ID.
9. Correlate the folder IDs obtained in step 4 with the 'getsyncfolders' API response entries in sync.log to obtain additional information pertinent to the shared folders (e.g., the fingerprints of the associated peers).
10. Examine the 'peers' subentry of the 'folders' entry (of the sync.dat file) to locate supporting information relating to the peers associated with the shared folders added by or downloaded to the device under investigation such as the last used IP addresses, last sync times, last seen times, and last data transfer times.
11. Correlate the device names obtained in step 7 with the log entries to identify the non-encoded peer IDs as well as identity-specific fingerprints.
12. Inspect the last modified time of each synced folder/file to determine if the files have been modified offline. If changes made are relevant to the investigation, remote synchronisation would assist in determining the changes made.
13. Once the peer IDs, device names, fingerprints, shared folders, share IDs, folder IDs, and other information of relevance are located, additional analysis can be undertaken on the OS-specific system and log files, web browsing information, carved files, memory files, network captures, and other evidence sources to provide supporting information or recover deleted data.
14. Iterate the framework with evidence source identification and preservation via the associated peer devices using the peer information (e.g., IP addresses) obtained.

**Presentation**

Merge the information gathered during the analysis phase into a report. This comprises:

1. The BitTorrent Sync version(s) installed on the device(s) investigated as well as the installation time(s).
2. The non-encoded peer IDs, device names, fingerprints, and IP addresses of the device(s) investigated as well as the associated peers.
3. The folder names, IDs, as well as timestamp information for the shared folders located in the device(s) investigated.
4. The peers which sent folder access requests to the device(s) investigated alongside the shared folder IDs, folder names, and timestamp information associated with the requests.
5. The peers associated with the share IDs or shared folders investigated.
6. Copies of the sync files as well as the timestamp information associated with the share IDs or shared folders investigated.

## 8. Conclusion and future work

Without the use of systematic investigative procedures and techniques, crucial evidence may be missed in an investigation and the integrity of the evidence would be compromised. In this paper, we investigated the artefacts from the use of BitTorrent Sync on computer and mobile devices running Window 8.1, Ubuntu 14.04.1 LTS, Mac OS X Mavericks 10.9.5, iOS 7.1.2, and Android KitKat 4.4.4. We then outlined an operational methodology with the view to provide a systematic approach for collecting and analysis of evidence relating to BitTorrent Sync.

Analyses of the directory listings revealed that the that the process of uploading, storage, and downloading files from/to the shared folders did not change the MD5 and SHA1 hash values, indicating the viability evidence collection through remote synchronisation. The synced files that were deleted could be forensically recovered from the peer device's 'Archive' subfolder as well as the non-emptied trash or recycle bin folder of the device which initiated the file deletion. In addition to directory listing, we were able to recover evidence of BitTorrent Sync usage from the log and metadata files such as settings.dat, sync.dat, history.dat, sync.log, and [share-ID].db located in the application folder. These files remained intact even after we unlinked the identity, with the exception of those located in the identity-specific */%BitTorrent Sync%/.SyncUser<Random number>* subfolder. Alternatively, it was identified that copies of the log and metadata files could be potentially recovered from memory files and unallocated space intact or in plain text, provided that the remnants are not cleared or flushed.

Analyses of the network captures determined that the network traffics were encrypted and the synced file remnants were not recovered. However, we were able to recover the peer discovery packets which contain information of forensic interest such the IP addresses, port numbers, peer IDs, and share IDs associated with the host and the peers.

Our examinations of the physical memory captures indicated that the memory dump can provide potential for alternative methods for recovering the login credentials and log and metadata files of forensic interest in plain text. The memory dump could also provide an alternative method for recovering the running process and network information using the '*pslist'* and 'netscan'/'netstat' functions of Volatility. The PIDs could assist the investigator in obtaining data associated with the Symform client application during further analysis of the physical memory dumps (i.e., locating the data remnants associated with the process using the *Yarascan* function of Volatility). The presence of the artefacts in the memory dump also means the artefacts could be potentially located in the swap files as a result of inactive memory pages being swapped out of the memory to the hard disk during the system's normal operation (Quick and Choo, 2013a, 2013b, 2014; Yang et al., 2016). Nevertheless, a practitioner must keep in mind that memory changes frequently according to users' activities and will be wiped as soon as the system is shut down. Hence, obtaining a memory snapshot of a compromised system as quickly as possible increases the likelihood of preserving the artefacts before being overwritten in memory.

A summary of findings from our research is shown in Table 11. To keep pace with technological advances, future work would include extending this research to other popular private cloud services (Syncany, Seafile, etc.) that can be used to inform IoT investigations.

Table 11: Summary of findings (R =Recoverable, P = Possibly Recoverable, N = Not Recoverable).

| **Platform** | **Source of Evidence** | **Data artefacts found** | | | | | | |
|---|---|---|---|---|---|---|---|---|
| | | **Installation, uninstallation, and BitTorrent Sync usage information** | **Peer device names/ IP addresses** | **Enron sample files/ path references** | **File-sharing Key** | **Share ID** | **Peer ID** | **Fingerprint** |
| **Windows 8.1** | **Directory listings** | R | R | R | R | R | R | R |
| | **Registry** | R | N | P | N | N | N | N |
| | **System logs** | R | N, only found in sync.log | N, only found in sync.log | N, only found in sync.log | N, only found in sync.log | N, only found in sync.log | N, only found in sync.log |
| | **Prefetch** | R | N | N | N | N | N | N |
| | **Thumbcache files** | P | N | P, synced image | N | N | N | N |
| | **Link files** | R | N | N | N | N | N | N |
| | **RAM** | P, including the log and metadata files | P | P | P | P | P | P |
| | **Pagefile.sys** | P, including the log and metadata files | P | P | P | P | P | P |
| | **Unallocated space** | P, including the log and metadata files | P | P | P | P | P | P |
| **Ubuntu 14.04.1 LTS** | **Directory listings/Stored files** | R | R | R | R | R | R | R |
| | **System logs** | N, only found in sync.log | N, only found in sync.log | N, only found in sync.log | N, only found in sync.log | N, only found in sync.log | N, only found in sync.log | N, only found in sync.log |
| | **Thumbcache files** | N | N | N | N | N | N | N |
| | **RAM** | P, including the log and metadata files | P | P | P | P | P | P |
| | **Swap partition** | P, including the log and metadata files | P | P | P | P | P | P |
| | **Unallocated space** | P, including the log and metadata files | P | P | P | P | P | P |
| **Mac OS X Mavericks 10.9.5** | **Directory listings/Stored files** | R | R | R | R | R | R | R |
| | **System logs** | R | N, only found in sync.log | N, only found in sync.log | N, only found in sync.log | N, only found in sync.log | N, only found in sync.log | N, only found in sync.log |
| | **Thumbcache files** | R | N | R | N | N | N | N |
| | **RAM** | P, including the log and metadata files | P | P | P | P | P | P |
| | **Swap partition** | P, including the log and metadata files | P | P | P | P | P | P |
| | **Unallocated space** | P, including the log and metadata files | P | P | P | P | P | P |
| **iOS 7.1.2** | **Directory listings/Stored files** | R | R | R | R | R | R | R |
| | **System logs** | R | N, only found in sync.log | N, only found in sync.log | N, only found in sync.log | N, only found in sync.log | N, only found in sync.log | N, only found in sync.log |
| **Android Kitkat 4.4.4** | **Directory listings/Stored files** | R | R | R | R | R | R | R |
| | **System logs** | N | N, only found in sync.log | N, only found in sync.log | N, only found in sync.log | N, only found in sync.log | N, only found in sync.log | N, only found in sync.log |
| **Network capture** | | R | R | N | R | R | R | N |